\documentclass[12pt]{article}

\usepackage[numbers,sort&compress]{natbib}
\usepackage{amsmath}
\usepackage{amssymb}
\usepackage{graphicx}
\usepackage{epstopdf}

\graphicspath{{fig/}}
\DeclareGraphicsExtensions{.eps}
\usepackage{epstopdf}

\newcommand{\Y}{{\mathcal Y}}
\newcommand{\C}{{\mathcal C}}
\newcommand{\bs}{{\bf s}}

\begin{document}
%
\title{Capacity-Achieving Rateless Polar Codes}
%
%
%
\author{Bin~Li,
        David~Tse,
        Kai~Chen,
        and Hui~Shen
\thanks{B. Li, K. Chen, and H. Shen are with the Communications Technology Research Lab., Huawei Technologies, Shenzhen, P. R. China (e-mail: {binli.binli,chenkai.chris,hshen@huawei.com}).}
\thanks{D. Tse is with the Department of Electrical Engineering, Stanford University and the Department of EECS, University of California, Berkeley  (e-mail: dntse@stanford.edu).}
}

%

\maketitle

\newtheorem{theorem}{Theorem}
\newtheorem{example}{Example}
\newtheorem{algorithm}{Algorithm}
\begin{abstract}

A rateless coding scheme transmits incrementally more and more coded bits over an unknown channel until all the information bits are decoded reliably by the receiver. We propose a new rateless coding scheme based on polar codes, and we show that this scheme is {\em capacity-achieving}, i.e. its information rate is as good as the best code specifically designed for the unknown channel.   Previous rateless coding schemes are designed for specific classes of channels such as AWGN channels, binary erasure channels, etc. but the proposed rateless coding scheme is capacity-achieving for broad classes of channels as long as they are ordered via degradation. Moreover, it inherits the conceptual and computational simplicity of polar codes.

\end{abstract}


\section{Introduction}

In  many communication scenarios, the quality of the communication channel is unknown to the transmitter. One possibility is to design a fixed-rate code for the worst possible channel but this often leads to an overly conservative solution. Another possibility is to use a {\em rateless code}, which transmits an increasing number of coded bits until all the information bits can be decoded reliably by the receiver. This solution requires the receiver to give simple ACK/NACK feedback to the transmitter but this capability is available in many communication scenarios. For example, rateless codes, appearing by the name of hybrid ARQ or incremental redundancy schemes,  are an essential part of a reliable and efficient wireless communication systems. As another example, rateless codes are very useful for packet erasure networks where the erasure rate is unknown.

A fixed-rate code designed for a specific channel is judged by its performance on that channel. In contrast, a rateless code is designed for a {\em class} of channels describing the channel uncertainty, and it is judged by its performance on {\em all}
 the channels in the class.  A rateless coding scheme is said to be {\em capacity-achieving} over a class of channels if for each channel in that class the number of coded bits transmitted by the scheme until reliable decoding is no more than the number of coded bits a capacity-achieving code specifically designed for that channel needs to transmit. Elementary information theoretic considerations show that random codes are capacity-achieving rateless codes for any class of channels which share the same capacity-achieving optimal distribution. A more interesting problem is the explicit construction of capacity-achieving rateless codes which have efficient encoding and decoding.  Two classes of such codes have been constructed.  First are the the LT codes
of Luby \cite{BLM02,Lub01} and the closely related  Raptor codes of Shokrollahi \cite{Sho06}. They are specifically designed for packet erasure channels. The second example is the rateless codes designed for AWGN channels by Erez, Trott and Wornell \cite{ETW12}. These rateless codes are built using fixed-rate capacity-achieving AWGN codes as base codes.

In this paper, we propose a new rateless coding scheme based on polar codes. Polar codes are the first class of low-complexity codes that are shown to achieve the capacity of a wide range of channels \cite{Arikan2009}. By leveraging this key property of polar codes, the rateless coding scheme we designed is shown to be  capacity-achieving for general classes of channels totally ordered by degradation. This is in contrast to the above two classes of rateless codes, each of which is designed for a {\em specific} such class (erasure channels and AWGN channels respectively).

One approach to designing a rateless coding scheme, used in turbo and LDPC based hybrid ARQ schemes, is {\em puncturing}: 1) a "mother" code with very low coding rate is first designed; 2) the mother code is significantly punctured and the remaining coded bits sent at the first transmission; 3) non-punctured bits are incrementally sent at later transmissions. But for polar codes, it is a problem to design a rateless scheme in this fashion. Due to the highly structured nature of polar codes, it is unclear how to to puncture a "mother" polar code with a low coding rate and maintain this punctured code as a "good" code. Nor is it clear how to incrementally add coding bits to a polar code with a very high coding rate and maintain the final low-rate code as a "good" polar code.

However, there is actually a very natural way to build a rateless scheme based on polar codes.  Recall that a fixed-rate polar code is constructed by applying a linear transformation recursively to convert the underlying channel into a set of noiseless channels and a set of completely noisy channels under successive decoding. Information bits are supposed to be transmitted on the noiseless channels while known (frozen) bits are transmitted on the completely noisy channels.  If the channel were  known at the transmitter, then the transmitter knows exactly which are the noiseless channels  and which are the completely noisy channel and this scheme can be implemented.  If the channel is unknown, then the transmitter does not know which channels are noiseless and which are completely noisy, but what it does know is a {\em reliability ordering} of the channels, such that regardless of what underlying channel is,  a more reliable channel is always noiseless if  a less reliable channel is noiseless, and a less reliable channel is always completely noisy if a more reliable channel is completely noisy.

Given this reliability ordering, a rateless scheme can be designed as follows. The initial transmission can be done aggressively using a high-rate polar code with many information bits and very few frozen bits. If this transmission cannot be decoded, then too many information bits are sent and too few bits are frozen. Among the information bits sent on the first transmission, the ones sent on the less reliable channels are retransmitted in future transmissions. By decoding these bits from the future transmissions, they effectively become frozen, allowing the rest of the information bits sent on the first transmission to be decoded. Thus, this scheme can be called {\em incremental freezing}, as future transmissions successively freeze more and more information bits sent in earlier transmissions.

In section \ref{sec_pre}, we present more details of the scheme and show that it is capacity-achieving. In section \ref{sec:sim}, we present a finite blocklength design with soft combining decoders and provide some simulation results. Finally we draw some conclusions.

\section{Rateless Polar Codes}
\label{sec_pre}

\subsection{Polar Codes Basics}

Given a binary-input channel $W$ with arbitrary output alphabet $\Y$, the first step of the standard polarization process creates two binary input channels:
\begin{eqnarray*}
W^-(y_1, y_2|x) &:= & \frac{1}{2} \sum_{u \in \{0,1\}} W(y_1|u +x) W(y_2|u) \\
W^+(y_1,y_2, u|x) &:=&  \frac{1}{2} W(y_1|u+x)W(y_2|x)
\end{eqnarray*}
This creates the $1$st-level channels. The channels at the $n+1$th level recursion can be constructed from the $n$th level channels. Given any length $n$ sequence $\bs$ of $+$'s and $-$'s, define the channels:
\begin{eqnarray*}
W^{\bs -} &: = &  (W^\bs)^-\\
W^{\bs +} & :=& (W^\bs)^+
\end{eqnarray*}
The theory of polarization \cite{Arikan2009} shows that as $n \rightarrow \infty$, a subset  ${\mathcal S}(W)$ of the $2^n$ $n$-level channels will have their mutual information converging to $1$, and the rest with mutual information converging to zero. By sending information bits on the former subset and "freezing" the latter subset with known bits, capacity can be achieved with a successive cancelation decoder. In the sequel, we will call ${\mathcal S}(W)$ the good bit indices. Note that this set depends on the original channel $W$.

\subsection{Degradedness and Nesting Property of Polar Codes}

A symmetric binary input channel $W_2$ is said to be {\em degraded} with respect to a channel $W_1$ if there exists random variables $X, Y, Z$ such that $X - Y - Z$ forms a Markov chain and the conditional distribution of $Y$ given $X$ is $W_1$ and the conditional distribution of $Z$ given $X$ is $W_2$.  We will use the notation $W_2 \preceq W_1$ and $W_1 \succeq W_2$.  For example, an AWGN channel of lower SNR is degraded with respect to an AWGN channel of higher SNR. An erasure channel of higher erasure probability is degraded with respect to an erasure channel of lower erasure probability, A BSC of higher crossover probability (less than 1/2) is  degraded with respect to a BSC of lower crossover probability.

It is known that the polarization operation preserves degradedness \cite{SW13}, i.e. if $ W_2 \preceq W_1$, then $W_2^+ \preceq W_1^+$ and $W_2^-  \preceq W_1^-$.  Following the recursion, this implies that $W_2^\bs \preceq W_1^\bs$ for any $\bs$. This  implies that in the limit of the polarization process, the good bit indices $\mathcal{S}(W_2)$  for $W_2$ (i.e. those with mutual information equal to $1$)  must be a subset of the good bit indices $\mathcal{S}(W_1)$ for $W_1$. We will call this the {\em nesting property}. This nesting property leads to the reliability ordering of the polarized channels mentioned in the introoduction.

\subsection{A Rateless Scheme: Basic Version}

We now present  a capacity-achieving rateless scheme built on polar codes.  We assume that communication is to take place over a class $\C$ in which the channels have binary-input and are symmetric, totally ordered via degradation and have capacities spanning a continuum from $0$ to $R$, where $R$ is called the peak-rate. We will show that the scheme is capacity-achieving in the following sense: For any integer $k \ge 1$,  if the capacity of the channel is between $R/(k+1)$ and $R/k$, then the scheme can achieve a rate of $R/(k+1)$ reliably. While the scheme is not truly rateless in the sense of achieving any {\em arbitrary} rate, a small modification of the scheme will make it rateless. This will be described in subsection \ref{sec:true_rateless}.


Consider a capacity-achieving  polar code of rate $R$  and (long)  block length $N$ designed for a channel $W_1$ whose capacity is $R$\footnote{While strictly speaking capacity-achieving  is a property of a sequence of codes of increasing block length, here we keep the language lightweight and discusses concepts in terms of  a code of fixed and long block length. Suitable limiting arguments can be made for a precise statement of the results.}. Let $\mathcal{S}(W_1)$ be the good bit indices.  Note that $|\mathcal{S}(W_1)| = NR$. At the first stage, we transmit all $NR$ information bits on $\mathcal{S}(W_1)$ and the rest of the bits are frozen. If the unknown channel $W$ is such that $W \succeq W_1$, then the receiver can decode after this transmission and we are done.

If the unknown channel $W$ is weaker than $W_1$, then the receiver cannot decode after the first transmission and the sender performs a second transmission. Let $W_2$ be the channel in $\mathcal{C}$ whose capacity is $R/2$. In the second transmission, we retransmit the information bits that were put on $\mathcal{S}(W_1) - \mathcal{S}(W_2)$ in the first transmission using the same polar code but with these information bits now put on $\mathcal{S}(W_2)$ and the rest of the bits frozen. Note that by the nesting property, $\mathcal{S}(W_2) \subset \mathcal{S}(W_1)$ so
$$|\mathcal{S}(W_1)-\mathcal{S}(W_2)| = NR/2 = |\mathcal{S}(W_2)|$$
and hence these bits can fit on $\mathcal{S}(W_2)$ on the second transmission. If $W \succeq W_2$, then the receiver can decode these bits based on the second transmission only. Using these bits as side information, the receiver now goes back to the first transmission and now the bits in $\mathcal{S}(W_1) - \mathcal{S}(W_2)$ becomes frozen bits and only the bits in $\mathcal{S}(W_2)$ need to be decoded. Since $W \succeq W_2$, these bits can be decoded as well based on the first transmission, and we are done.

If the unknown channel $W$ is weaker than $W_2$, then the receiver cannot decode after the first transmission and the sender performs a third transmission. Let $W_3$ be the channel in $\mathcal{C}$ whose capacity is $R/3$. If the unknown channel were $W_3$, then the bits sent on $\mathcal{S}(W_2)- \mathcal{S}(W_3)$ in both first and second transmissions should have been frozen, but they were not. So the sender re-transmit these information bits in the third transmission. The number of such information bits are
$$ 2(NR/2 - NR/3) = NR/3,$$
so they can all be transmitted on the $\mathcal{S}(W_3)$ indices in the third transmission (with the rest of the bits frozen). If the unknown channel $W$ is equal or stronger than $W_3$, then all these bits can be correctly decoded from the third transmission. Among these bits, the ones sent on the second transmission are now side information to be used to decode {\em all} the information bits sent on the second transmission. These latter bits, together with the bits that are decoded from the bits that are decoded from the third transmission and are repeated directly from the first transmission, become side information to freeze the bits sent on $\mathcal{S}(W_1) - \mathcal{S}(W_3)$ indices in the first transmission, enabling the bits sent on $\mathcal{S}(W_1)$ in the first transmission to be decoded as well.

In general, suppose after $k$ transmissions, decoding has failed. Now we shoot for a channel $W_{k+1}$ of rate $\frac{R}{k+1}$. We retransmit all the information bits sent on $\mathcal{S}(W_{k}) - \mathcal{S}(W_{k+1})$ in all the previous $k$ transmissions. There are a total of
$$ k \left (\frac{NR}{k} - \frac{NR}{k+1} \right)=  \frac{NR}{k+1}$$
such bits, and they can all be sent on the $\mathcal{S}(W_{k+1})$ indices in the $k+1$th transmission. Using the backward decoding strategy described above, we can now go back and decode everything.

\begin{figure}[!b]
  \centering
  \includegraphics[width=0.95 \columnwidth]{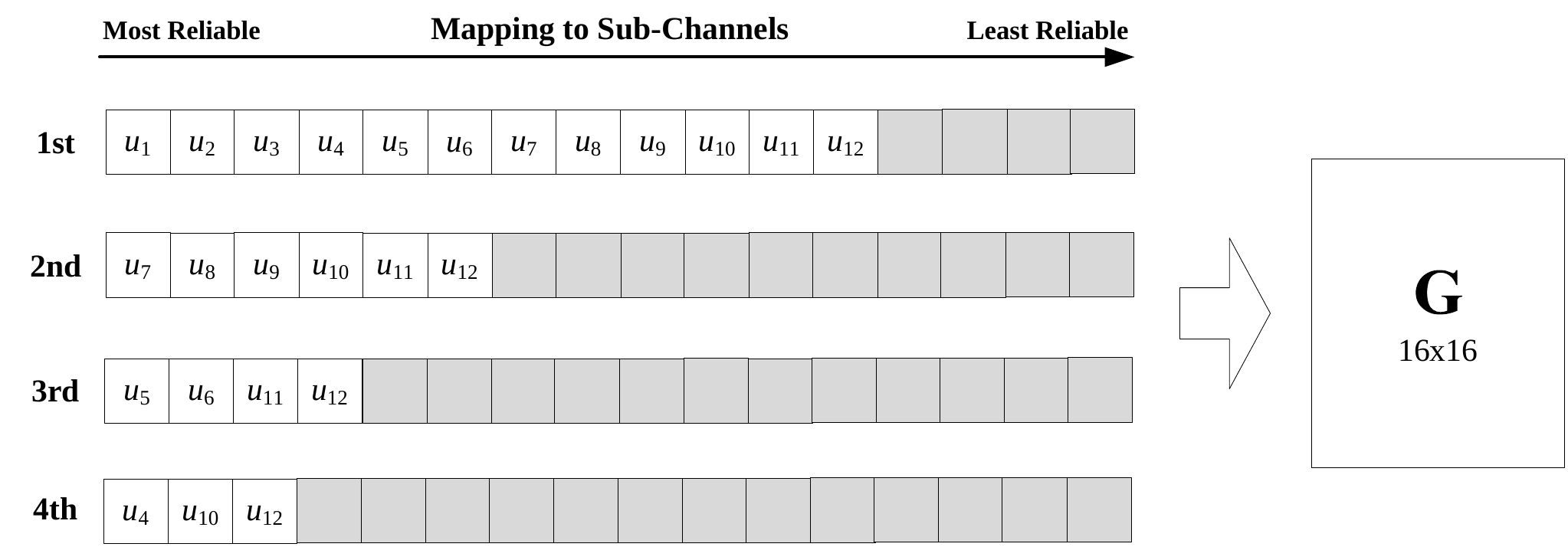}
  \caption{A simple example of incremental freezing with $N=16$, $K=12$, and up to $4$ transmissions.}
  \label{fig_eg}
\end{figure}

\subsection{An Example}

Figure  \ref{fig_eg} gives a simple example with $N=16$, $K=12$ for the peak-rate code, i.e. we shoot for a maximum rate $R = 3/4$.

In the initial transmission, the $12$ information bits $u_1, u_2, \cdots, u_{12}$ are sent on the $12$ channels with largest reliabilities, the initial rate $R_1 = R = 3/4$. Other bits are frozen.

When the 1st transmission fails, the second half of the information bits, $u_7, u_8, \cdots, u_{12}$ are sent on the $6$ most reliable channels of the 2nd transmission; the rate for the first and second transmissions combined is $R_2=6/16=R/2$.

When the 2nd transmission also fails, $u_5, u_6$ from the first transmission and $u_{11}, u_{12}$ from the second transmission are sent on the best $4$ channels of  the 3nd transmission; the rate for the first three transmissions combined  is $R_3=3/16=R/3$.

Finally, when the 3rd transmission still fails, $u_4, u_{10}, u_{12}$ are sent on the best $3$ channels for the 4th transmission; the rate for the first four transmissions combined  is $R_4=3/16=R/4$.

\subsection{Incremental Freezing}

We can think of the above scheme  as {\em incremental freezing} of information bits in a polar code. If we knew the channel, then we would freeze exactly the right bits. However, since we don't know what the channel is, we should be more aggressive and freeze few bits and send many information bits. If we are lucky and the channel is good, then all the information bits get through. On the other hand, if the channel is not as good as we hope, then we can retroactively freeze more bits by retransmitting and decoding these bits in future transmissions. Because of the nesting property, we know exactly what bits we should retroactively freeze. And we effectively freeze more and more bits incrementally as the the actual channel is worse and worse than expected.

The key to why the scheme is capacity-achieving is indeed the nesting property. At each transmission, we don't know what the unknown channel $W$  is, but we are always assured that the good bit indices for the unknown channel is a subset of the information bits used in that transmission. So we never "waste" any mutual information in any transmission. We may not be able to decode the  bits because too few bits are frozen, but this can always be rectified by retroactively freezing.

\subsection{Extension to Arbitrary Rates}
\label{sec:true_rateless}

The above scheme is not truly rateless as it can only achieve the rates $R$,$R/2, R/3, \ldots, $ rather than a set of arbitrary rates. A simple extension of the scheme will rectify this issue. The idea is that in future transmission, {\em new} information bits can be transmitted in combination with  bits retransmitted from previous transmissions. For example, suppose we want to achieve rates $R$ on the first transmission, and $R - \Delta$ bits on the first and second transmission combined, where $\Delta > 0$ is arbitrary.  The first transmission is exactly the same as before. In the second transmission, instead of re-transmitting the information bits sent on the least $NR/2$ reliable positions in the first transmission, one should retransmit only the bits sent on the least  $N\Delta$ reliable positions in the first transmission and add $N(R - 2\Delta)$  new information bits to send a total of $NR$ bits in the most reliable positions of the second transmission.

\subsection{Extension to Parallel Channels}
\label{sec:parallel}

It is known that polarization holds for much more general channels than  symmetric binary-input channels \cite{STA09}. So presumably one can extend the theory of rateless polar coding to more general classes of channels beyond binary symmetric ones. Here we purse one such generalization: parallel channels.

A parallel channel W is composed of $Q$ independent component sub-channels $W^{(1)}, \ldots W^{(Q)}$. Here, we focus on parallel channels whose component channels have binary inputs and are symmetric.  We are interested in such parallel channels because they model AWGN channels with $2^Q$- PAM input. Using techniques like bit interleaved coded modulation (BICM), the AWGN channel with $2^Q$-PAM input can be modeled as a parallel channel with $Q$ binary input AWGN channels.

 A parallel channel W is degraded with respect to a channel $V$ if each of the components of $W$ is degraded with respect to the corresponding  component of $V$.  Let $\C$ be  a class of  parallel channels which are totally ordered via degradation. This can for example model a class of AWGN channels with $2^Q$-PAM input, where all the component sub-channels depend on the (single) SNR of the AWGN channel. We now show that  our rateless scheme extends naturally to this class of channels.

Consider a parallel channel  $W_1$ with capacity $R$; we decompose it into $Q$ parallel sub-channels with rates: $R_{11}, R_{12},\cdots,R_{1Q}$, where $R=R_{11}+R_{12}+\cdots+R_{1Q}$. For the first transmission, we design $Q$  polar codes of block length $N$  for the $Q$ parallel sub-channels and rates: $R_{11}, R_{12},\cdots,R_{1Q}$, respectively. If the unknown channel $W$ is such that $W \succeq W_1$, then the receiver can decode after this transmission and we are done.

If the unknown channel $W$ is weaker than $W_1$, then the receiver cannot decode after the first transmission and the sender performs a second transmission. Let the capacity of $W_2$ be $R/2$ and the capacity of the $Q$ parallel channels be $R_{21}, R_{22},\cdots,R_{2Q}$, where $R/2=R_{21}+R_{22}+\cdots+R_{2Q}$. In the second transmission, we retransmit $NR/2$ information bits. These bits are $NR_{11}-NR_{21}/2$ bits from the information bits of the  polar code used for the first sub-channel, $NR_{12}-NR_{22}/2$ bits from the information bits of the polar code used for the second sub-channel,$\cdots$, and $NR_{1Q}-NR_{2Q}/2$ bits from the information bits of the polar code of the $Q$ th sub-channel. These bits are then distributed into the $Q$ Polar codes with rates: $R_{21}, R_{22},\cdots,R_{2Q}$, respectively. 
If $W \succeq W_2$, then the receiver can decode these bits based on the second transmission only. Using these bits as side information, the receiver now goes back to the first transmission.  Since $W \succeq W_2$, the first transmission can be decoded as well.

If the unknown channel $W$ is weaker than $W_2$, then the receiver cannot decode after the first transmission and the sender performs a third transmission. Let the capacity of $W_2$ be $R/3$ and the capacitiesof the $Q$ sub-channels be $R_{31}, R_{32},\cdots, R_{3Q}$, where $R/3=R_{31}+R_{32}+\cdots+R_{3Q}$. In the third transmission, we retransmit $NR/3$ information bits. These bits come from the information bits of $Q$ polar codes used in the first transmission and from the $Q$ Polar codes used in the second transmission. More precisely, there are $NR_{21}-NR_{31}/2$ bits from the information bits of the two Polar codes for the first and second transmission for the first parallel channel, $NR_{22}-NR_{32}/2$ bits from the information bits of the two Polar codes for the first and second transmission for the second parallel channel,..., $NR_{2Q}-NR_{3Q}/2$ bits from the information bits of the two Polar codes for the first and second transmission for the $Q$th parallel channel. These bits are then distributed into $Q$ Polar codes with rates: $R_{31}, R_{32},\cdots, R_{3Q}$, respectively.

In general, the $k$th transmission sends $NR/k$ information bits. They are collected from $NR/k/(k-1)$ information bits sent at each previous transmission. To decode the $m$th transmission ($1 \le m \le k$), the receiver uses the side information from $(m+1)-$th, $(m+2)$-th, $\cdots$, $k$th decoded data to freeze
$$ RN \left (\frac{1}{m(m+1)} + \frac{1}{(m+1)(m+2)} + \cdots + \frac{1}{(k-1)k} \right) $$
$$=  RN \left (\frac{1}{m} - \frac{1}{k} \right) $$
information bits and only need to decode $$ \frac{RN}{m} - RN \left (\frac{1}{m} -  \frac{1}{k} \right)=  RN \left (\frac{1}{m} \right) $$ information bits. Note that with the side information, all transmissions have the same rate $R/k$ after $k$ transmissions.

\begin{figure}[!b]

  \centering
  \includegraphics[width=0.75 \columnwidth]{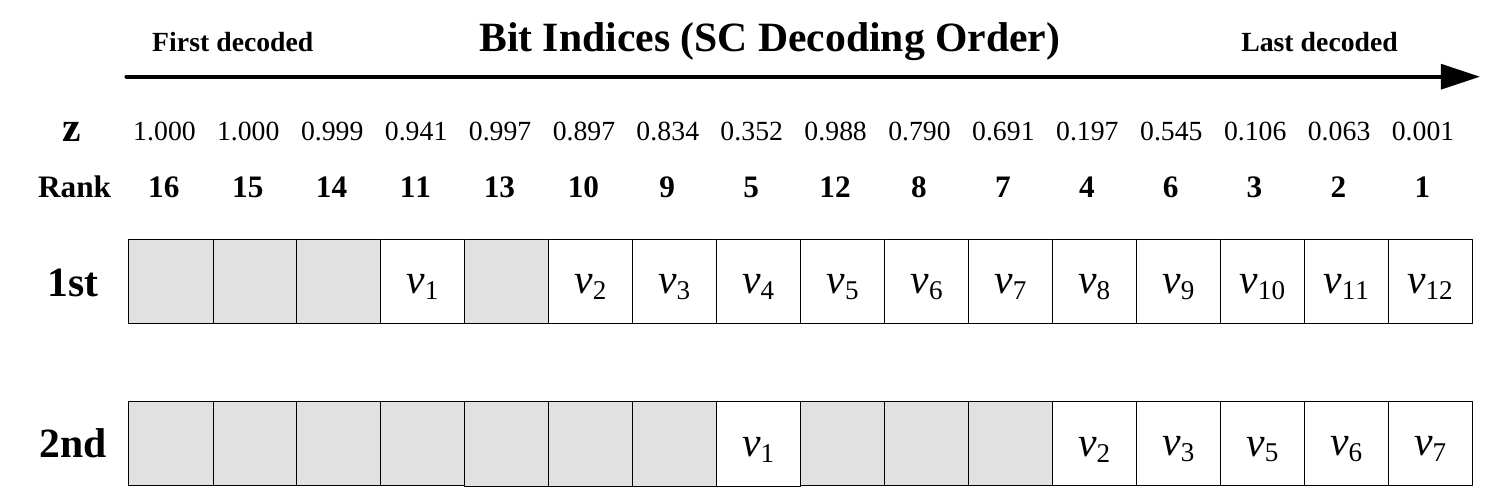}

  \includegraphics[width=0.75 \columnwidth]{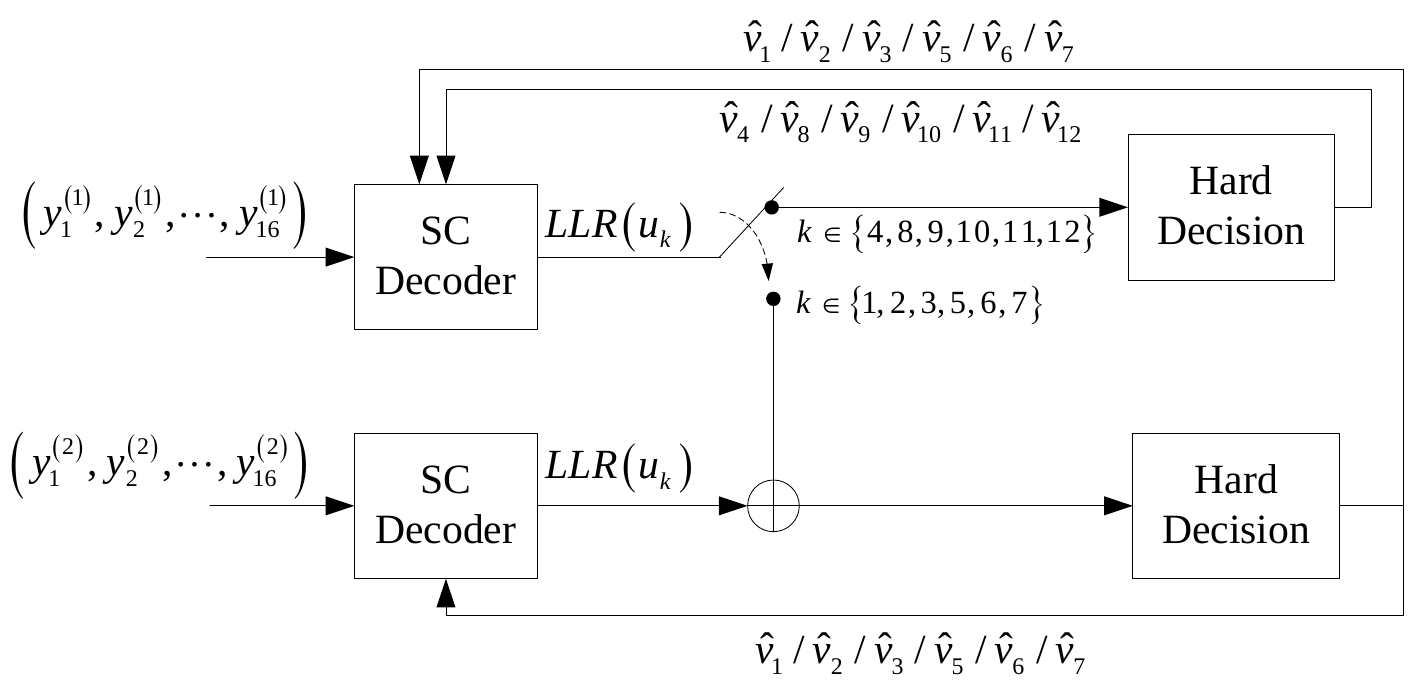}
  \caption{Upper figure: Transmission. Note that unlike Figure \ref{fig_eg}, the channels here are ordered by the successive cancelation order rather than by reliability order.
Lower figure:  Soft decoding. Bits like $v_1$ that are transmitted twice are estimated by combining the LLR's from both transmissions.}
  \label{fig:soft}
\end{figure}

\section{Simulation Results}
\label{sec:sim}

We performed some simulations over binary-input AWGN channels to assess the finite block length performance of the rateless polar coding scheme we proposed.

In the scheme proposed, we assume that retransmitted bits are decoded based only on the received signal of the retransmission. This is sufficient to achieve capacity. However, reception from the previous transmissions of the bits still provide some information, and finite block length performance can be improved by using a soft decoder that combines the reception across multiple transmissions. We have designed such a soft decoder. An example of how it operates can be seen in Figure \ref{fig:soft}.

\begin{figure}[!b]
  \centering
  \includegraphics[width=0.75 \columnwidth]{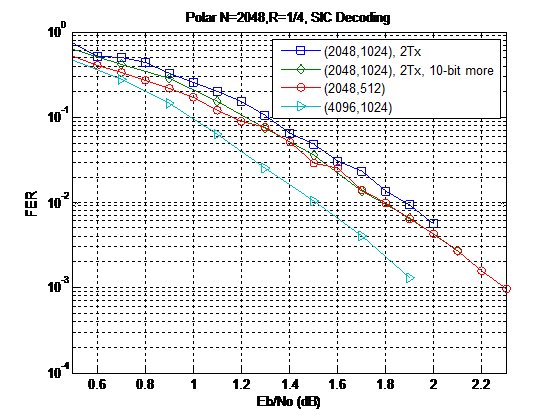}
  \caption{SC performance.}
  \label{fig:sc_performance}
\end{figure}

\begin{figure}[!b]
  \centering
  \includegraphics[width=0.75 \columnwidth]{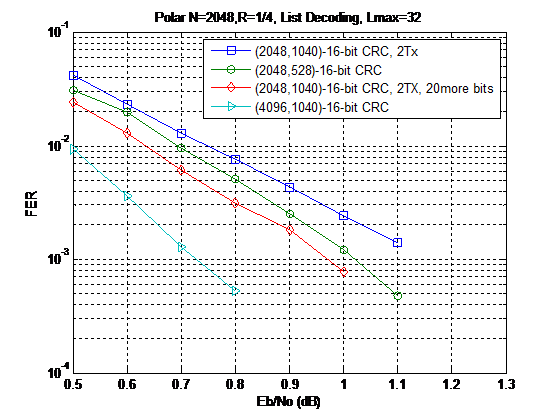}
  \caption{List Decoder performance.}
  \label{fig:list_performance}
\end{figure}

Figure \ref{fig:sc_performance} shows the  performance of the soft-combining SC decoder. The peak-rate polar code is $(2048. 1024)$, yielding a peak rate of 1/2. We evaluate the frame error rate after the second transmission, i.e. effective rate is $1/4$. The blue curve shows the performance of the scheme. In the scheme, the number of bits from the first transmission retransmitted in the second transmission is exactly the same as the number of bits not retransmitted. But actually the retransmitted bits get slightly better treatment since they can be decoded by combining receptions from both transmissions. To take advantage of this, a few more bits can be retransmitted. By optimizing this number, a small improvement can be obtained. This is shown by the green curve, where $10$ more bits are retransmitted. This optimized performance essentially matches that of a block length $2048$, rate $1/4$ code (red curve), but is about $0.3$ dB from a block length $4092$ polar code of rate $1/4$ designed for that signal-to-noise ratio. Note that since the effective block length of the rateless scheme after two transmissions is $4092$, we see that there is still a gap, albeit small,  with a code optimized for that SNR. This is the price of being rateless.

Figure \ref{fig:list_performance} shows the performance using a  soft-combining adaptive SC-list decoder with CRC \cite{Tal1206,Li2012}. We can see that an optimized rateless scheme is about $0.25$ dB from a polar code with block length $4096$

\section{Conclusion}
\label{sec_con}

In this paper, we propose a capacity-achieving rateless scheme based on polar codes. At the first transmission, a  peak-rate polar code is used, and at the later transmissions, information bits are retransmitted and decoded, and hence  incrementally  freezed to allow for decoding at the first transmission. The conceptual simplicity of the scheme attests to the inherent flexibility of polar codes.

\end{document}